\documentclass[twocolumn]{aastex62}

\usepackage{comment}
\usepackage[normalem]{ulem}

\urldef\techdocs\url{https://acdc-hst.readthedocs.io/en/latest/about.html#the-finer-details}

\usepackage[graphicx]{realboxes} 
\accepted{\today}
\submitjournal{PASP}

\shorttitle{COS Background Correction}
\shortauthors{Hernandez et al.}

\begin{document}

\title{Pushing the limits of the Cosmic Origin Spectrograph (COS) with an optimized background correction}

\correspondingauthor{Svea Hernandez}
\email{sveash@stsci.edu}

\author[0000-0003-4857-8699]{Svea Hernandez}
\affiliation{AURA for ESA, Space Telescope Science Institute, 3700 San Martin Drive, Baltimore, MD 21218, USA}

\author[0000-0003-2145-1022]{Andrei Igoshev}
\affiliation{Department of Applied Mathematics, University of Leeds, Leeds LS2 9JT, UK}

\author[0000-0003-4068-5545]{Jo Taylor}
\affiliation{Space Telescope Science Institute, 3700 San Martin Drive, Baltimore, MD 21218, USA}

\author{David Sahnow}
\affiliation{Space Telescope Science Institute, 3700 San Martin Drive, Baltimore, MD 21218, USA}

\author{Logan Jones}
\affiliation{Space Telescope Science Institute,
3700 San Martin Drive, 
Baltimore, MD 21218, USA}

\begin{abstract}
Observations utilizing the ultraviolet capabilities of the Cosmic Origin Spectrograph (COS) onboard the Hubble Space Telescope are of unique value to the astronomy community. Spectroscopy down to 900 \r{A} with COS has enabled new science areas. However, contrary to the situation at longer wavelengths, these observations are limited by detector background noise. The background correction currently applied by the standard calibration pipeline (\texttt{CalCOS}) is not optimized for faint targets, limiting the scientific value of low signal-to-noise observations. In this work we investigate a possible dependence of the variations of the dark rate in both segments of the COS far-ultraviolet (FUV) detector on time, detector high voltage (HV), and solar activity. Through our analysis we identified a number of detector states (on a configuration basis, e.g., HV and segment) characterizing the spatial distribution of dark counts, and created superdarks to be used in an optimized 2-dimensional (2D) background correction. We have developed and tested Another COS Dark Correction (\texttt{ACDC}), a dedicated pipeline to perform a 2D background correction based on statistical methods, producing background-corrected and flux-calibrated spectra. While our testing of ACDC showed an average improvement in S/N values of $\sim 10$\%, in a few cases the improvements in S/N reached 60\% across the whole wavelength range of individual segments.
 \end{abstract}


\section{Introduction}\label{sec:intro}
The Space Telescope Imaging Spectrograph (STIS) and the Cosmic Origins Spectrograph (COS) onboard the Hubble Space Telescope (HST) have demonstrated the significant impact of ultraviolet (UV) observations on modern astronomy. Until the full-scale development of the next generation UV/Optical/Infrared observatories, e.g., Habitable Worlds Observatory\footnote{\href{https://science.nasa.gov/astrophysics/programs/habitable-worlds-observatory/}{https://science.nasa.gov/astrophysics/programs/habitable-worlds-observatory/}}, there are no prospects for large space missions with this integral coverage of the electromagnetic spectrum.  Therefore, observations taken with these two instruments are of legacy value. We must continue to push the limits of these HST spectrographs, not only with future observations, but also through the usage of archival data. \par
At wavelengths $<$1130 \r{A} the sensitivity of COS declines by a factor of 100 (compared to that at higher wavelengths), and is comparable to that of FUSE \citep{mcc10,hir23}. As a result, the COS coverage of such blue wavelengths has played an important role in scientific studies involving He II re-ionization \citep{wor11,wor16}, Lyman continuum escape fraction from low-z galaxies \citep{bor14, izo16,izo16b,izo18,izo18b,izo21, lei16,her18, flu21a, flu21b}, circumgalactic medium \citep{tum13, mcc21}, and cold ISM \citep{jam14, her20}, to name a few. For many of these studies, observations are taken in the background-limited regime where an optimal detector background subtraction is critical. One example is the starburst galaxy Mrk 54 observed as part of HST program ID 13325, with a count rate at $\sim$910 \r{A} (rest frame) of $\sim$1.3$\times$10$^{-6}$ counts/sec/pix, compared to the dark rate of COS at the time, $\sim$4.8$\times$10$^{-6}$ counts/sec/pix. \par

\begin{figure*}
   
   	  \centerline{\includegraphics[scale=0.6]{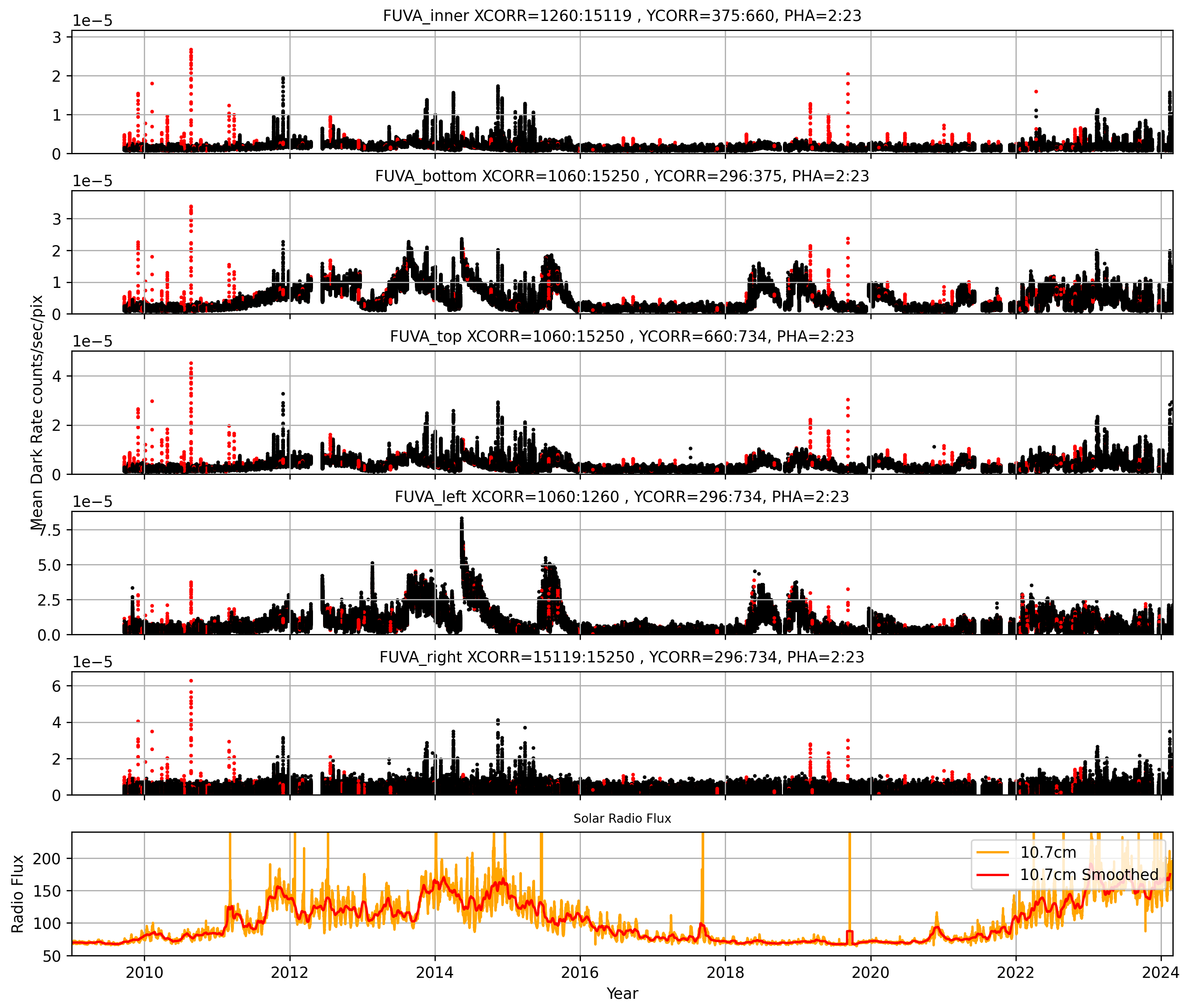}}
      \caption{ COS FUV (segment A) dark rate as a function of time as measured in the COS FUV Detector Dark Monitor\textsuperscript{a}. Each panel represents a different region on the detector. The dark rate estimates are obtained from COS dark exposures stored in the MAST archive. The red points show dark rate values observed close to the South Atlantic Anomaly. The trends in the different panels highlight how the dark rate varies spatially. The bottom panel shows the  solar radio emission at 10.7 cm (yellow line, in units of 10$^{-22}$ W m$^{-2}$ Hz$^{-1}$) by the National Oceanic and Atmospheric Administration, NOAA\textsuperscript{b}, with a smoothed version overplotted (red line).}
      \small\textsuperscript{a} \href{https://www.stsci.edu/hst/instrumentation/cos/performance/monitoring}{https://www.stsci.edu/hst/instrumentation/cos/performance/monitoring}\par
      \small\textsuperscript{b} \href{https://www.swpc.noaa.gov/products/solar-cycle-progression}{https://www.swpc.noaa.gov/products/solar-cycle-progression}
     
            \label{fig:monitor}
   \end{figure*}

We refer to counts not originating from photons incident on the detector as \textit{dark counts}. These counts are a combination of radioactive decay of atoms in the microchannel plates; charged particles in the environment of HST; and other sources internal to the instrument. It has been previously reported that the COS FUV background is somewhat correlated with Solar activity \citep{das19}, detector gain, high voltage (HV), and time (Figure \ref{fig:monitor}). The exact behavior of these background variations is not yet fully understood. The first five panels in Figure \ref{fig:monitor} clearly highlight the variations in dark rate depending on the region of the detector and the time of the observations. Given the complex dependencies of the COS background with each of these parameters, a robust and well-characterized background correction as a function of time is required for an accurate analysis of observations of faint FUV targets.\par
The standard COS calibration pipeline, \texttt{CalCOS}\footnote{\href{https://hst-docs.stsci.edu/cosdhb/chapter-3-cos-calibration/3-3-CalCOS-structure-and-data-flow}{https://hst-docs.stsci.edu/cosdhb/}}, determines the background contribution to science spectra by computing the average counts in pre-defined regions in that particular exposure. For COS FUV observations there are two pre-defined background regions (typically below and above the science extraction region). The dark rate at the target location can differ from that at the pre-defined regions used for computing the background contributions to the target spectrum.
Additionally, there are pixel-to-pixel variations in the spectral extraction region possibly differing from those at the background regions, which can lead to the over- or under-subtraction of the dark counts in the final spectrum.
 Performing a robust Pulse Height Amplitude (PHA; see Section \ref{sec:char}) screening \citep{ely14} along with optimizing the size of the extraction box can help reduce the background contribution. However, these practices do not guarantee an accurate background calibration. In addition to these methods, and given the importance of a reliable background subtraction, different research collaborations have independently worked to improve the calibration of low signal-to-noise (S/N) FUV observations \citep[e.g., ][\texttt{FaintCOS}]{wor16,lei16, mak21}. \par

The COS FUV G140L/800 setting (central wavelength at 800 \r{A}) was made available to the community in 2018, and was primarily designed for background-limited  observations \citep{red16}. This new configuration decreases the background levels by a factor of two, and increases the S/N of background-limited datasets, with the caveat that it is restricted to low resolution observations ($R\sim$2000). To further optimize the calibration of COS data products and especially benefit users observing in the limits of Poisson noise, here we present recently developed software to improve low S/N ($\lesssim$5) COS FUV observations by applying a more accurate and tailored background correction. This new background correction is performed on the 2-dimensional (2D) COS science exposures, accurately modeling and subtracting the expected dark counts in the science extraction region itself. \par
This paper is organized as follows. In Section \ref{sec:obs} we describe the COS dark observations used to characterize the background spatial and temporal variations. We present a brief summary of our in-depth investigation on the COS background noise in Section \ref{sec:char}. Section \ref{sec:correction} provides an overview of the parametric background correction and the adopted data-driven approach. And lastly, in Sections \ref{sec:acdc} and \ref{sec:conclusion} we introduce our newly-developed software \texttt{ACDC} and present our concluding remarks, respectively. We note that a much more detailed technical report describing our investigation into the background properties of the instrument can be found in our official documentation page\footnote{\href{https://github.com/jotaylor/acdc-hst}{https://github.com/jotaylor/acdc-hst}}.


\section{Description of Observations}\label{sec:obs}
The spatial characterization of the behaviour of the dark counts on the COS FUV detector relied on the compilation of all the FUV dark exposures stored in the STScI archive, Mikulski Archive for Space Telescopes (MAST\footnote{\href{https://mast.stsci.edu}{https://mast.stsci.edu}}). We retrieved from MAST a total of 7190 COS FUV dark exposures (including both segments A and B) available as of September 2023. The observations cover a time period between June 06, 2009 and September 11, 2023. The majority of the dark exposures were taken as part of COS FUV detector dark monitoring programs, which nominally collect five individual exposures every week, each with an exposure time of 22 minutes (1330 seconds). This set of 7190 exposures were taken at different HVs, with the most widely used being HV = 163, 167, 169, 171, 173, and 178 digital units (du). As detailed in \citet{das19}, the conversion between du and volts follows \begin{equation} HV = -(du \times 15.69 + 2500)\end{equation} \par

To facilitate the analysis and characterization of the COS background, once the dark exposures were retrieved from the archive, these were ingested into (1) an \texttt{SQLite} database, and (2) a \texttt{MySQL} database, each with different formats. The \texttt{SQLite} database was designed to track the dark counts on different regions of the FUV detector over 25-second intervals. For each dark exposure, the dark counts are measured in five separate and standard regions of the detector \citep[comparable to those defined in Figure \ref{fig:monitor}; see Figure 9 in ][]{das19}, and inserted into the \texttt{SQLite} database. The parameters used to define these regions, as adopted by the COS team, are detailed in Table \ref{tab:cos_regions}. 
The second database uses \texttt{MySQL}, and rather than tracking dark counts in time segments, each row stores information for each individual dark event recorded in the suite of COS dark exposures. The database contains 8 tables, one for each of the typically used HV settings (163, 167, 169, 171, 173, 175, 178) and one for all other HVs, often used in experimental or calibration observations. 

\begin{table}[!h] 
  \centering
    \caption{COS detector regions used in the monitoring of the dark rates.}\label{tab:cos_regions}
    \def\arraystretch{1.25}
     \begin{tabular}{l c c c c}
    \hline
    \hline
   Region & Min $x$ & Max $x$ & Min $y$ & Max $y$ \\
    \hline
    \multicolumn{5}{c}{Segment A} \\
    \hline
    Inner & 1260   & 15119 & 375  & 660 \\
    Left& 1060  &  1260  & 296  & 734 \\
    Right & 15119 &  15250  & 296 & 734 \\
    Bottom & 1060 &  15250  & 296 & 375 \\
    Top & 1060 &  15250  & 660 & 734  \\
    \hline
    \multicolumn{5}{c}{Segment B} \\
    \hline
    Inner & 1000   & 14990 & 405  & 740 \\
    Left& 809  &  1000  & 360  & 785 \\
    Right & 14990 &  15182  & 360 & 785 \\
    Bottom & 809 &  15182  & 360 & 405 \\
    Top & 809 &  15182  & 740 & 785  \\
    \hline
    \end{tabular}
\end{table}

\section{Characterization of the COS background}\label{sec:char}
As part of this work we preformed an in-depth investigation on the factors influencing the background levels both globally and spatially, with a special focus on the pulse height amplitude (PHA) from each event. In brief, for each photon that lands on the COS FUV detectors, a cascade of electrons is created which is characterized by a PHA that is then registered by the electronics (see Section 4.1.7 in \citealt{hir23}). It has been generally understood that external events are distinguished from background noise events by their PHA values, where dark events typically have low PHAs (PHA = 0 - 2) and real/external events register higher PHA values (PHA $\geq$ 3). As noted in \citet{sah11}, background counts are expected to display a negative exponential pulse height distribution (PHD). In contrast, the PHD observed for photon events is more commonly quasi-gaussian.\par 
For a detailed description of our work we refer to the technical document in our public page\footnote{\techdocs}. Briefly summarized, we found that the PHDs from dark counts do not necessarily follow a simple power-law trend, instead they show a more complex distribution. While dark counts do show some correlation with the solar flux, this correlation is weak and only statistically significant in segment A. Additionally, we found that the spatial structure and temporal distribution of the dark counts are neither easily nor accurately described by a simple model. Therefore, our approach for modeling the dark counts is primarily data-driven.\par

\section{Parametric background correction}\label{sec:correction}
The issue of accurately correcting the background contributions to the science spectrum can be described as follows: in a given  pixel we observe a total number of counts defined as $N = n_s + n_d$ in each science exposure with exposure time of $t_s$. This number consists of $n_s$ counts from the astronomical source and $n_d$ counts from other sources, already identified as dark counts. We assume that both of these counts follow a Poisson distribution $n_s \sim \mathrm{Pois} (\psi_s)$ and $n_d \sim \mathrm{Pois} (\psi_d)$ where $\psi_s$ and $\psi_d$ are the true fluxes for the target signal and dark counts, respectively. It is important to note that the total number of counts, $N$, also follows a Poisson distribution, namely $N\sim \mathrm{Pois} (\psi_s + \psi_d)$. In a general case, the probability mass function for the Poisson distribution is described as
\begin{equation}
f(k, \psi) = \frac{\psi^{k}e^{-\psi}}{k!}    
\end{equation}
where $k$ is the total number of events, in this case $N$. \par
In a generic COS FUV observation, it is likely that the scientific source is not precisely well-characterized, i.e., $n_s$. Instead, our approach is to analyze detector regions outside of both the science extraction region and the Wavelength Calibration Aperture (WCA) region, where the registered counts consist only of dark signal, therefore, $N=n_d$. Using this approach, we focus on accurately modeling the 2D spatial dark variations by using a combination of empirical superdarks. This allows us to parameterize the background subtraction, fixing the value $\psi_d$ using either the empirical superdarks or background model.
We adopt a data-driven modeling approach where we assume a finite number of detector states describing the dark behavior: $Q$, $P_1$, ... $P_n$, where $Q$ is a quiescent state and $P_1$, ..., $P_n$ are peculiar states of the detector. Furthermore, we assume that at any given time the dark count distribution can be described as a linear combination of a quiescent and peculiar states:
\begin{equation}
W = \alpha_1 Q + \sum_{i=1}^N \alpha_{i+1} P_i    
\end{equation}
The optimal values for the coefficients $\alpha_i$ are obtained through likelihood maximization. This maximization approach is done through the analysis of the 2D science exposures, intentionally excluding the counts registered in the science extraction and WCA regions. In an effort to obtain the best background model, our optimized software iterates over each pixel to estimate the dark count distribution, $W(x,y)$, and minimizes C statistic \citep{Cash1979ApJ} as follows:
\begin{equation}
C = 2 \sum_{i=0}^N W_i  - N_i \log (W_i) + \log (N_i!)    
\end{equation}
where $W_i$ is the predicted number of dark counts in a pixel with index $i$ and $N_i$ is the recorded number of photons in the same pixel during the science exposure.

\begin{table*}
\caption{Low S/N testing suite}
\label{tab:test_suite}
\centering 
\begin{tabular}{ccccccccc}
\hline \hline
   Target & PID & Dataset & Date & Grating & Cenwave & Median S/N & HV$_{\rm SegA}$ & HV$_{\rm SegB}$\\
\hline
KISSR1637&	11522&	LB6206010&	10/24/10&	G130M&	1291	&2	&169&	167\\
SDSS-J145735.13+223201.8&	13293&	LCAG02010&	8/29/13&	G160M&	1577&	2&	167&	169\\
IC-1586&	13481&	LCDR01010&	11/28/13&	G140L&	1105&	2&	167&	-\\
HCG92-7&	13321&	LCAY06020&	8/13/14&	G130M&	1222&	1&	167&	175\\
J1152+3400&	13744&	LCM802030&	5/8/15&	G140L&	1280&	1&	167&	163\\
J1333+6246&	13744&	LCM803030&	7/6/15&	G140L&	1105&	1&	167&	-\\
SDSS-J103020.91+611549.3&	13654&	LCOX08010&	9/12/15	&G130M&	1327&	2&	167&	163\\
GP1205+2620&	14201&	LCXR16010&	2/28/16&	G160M&	1577&	1&	167&	169\\
J1107+4528&	14079&	LCTD19010&	01/23/17&	G130M&	1291&	1&	167&	175\\
2MASS-J15570234-1950419	&15310&	LDMP24010&	8/2/18&	G130M	&1222	&1&	163&	167\\
J081112+414146&	15626&	LDXE11010&	9/17/19&	G140L &	800&	1	&163&	-\\
J003601+003307&	15626&	LDXE08010&	9/25/19&	G140L&	800&	1&	163&	-\\
J124423+021540&15626&	LDXE43010&	3/25/20&	G140L& 	800&	1	&163&	-\\
SDSS154714.35+175153.1&	17115&	LF1411010&	4/29/23&	G130M&	1291&	1&	167&	169\\
NGC-1313-P2&	17180&	LEVH02010&	5/31/23&	G130M&	1291	&2	&167	&169\\
BD-10-47&	16701&	LEOO11010&	6/18/23	&G130M	&1222	&1&	173	&175\\
\hline
\end{tabular}
\end{table*}

\subsection{Background Models: Multi-states}\label{sec:multi_states}
Given the complex correlations between the spatial and temporal trends with respect to solar activity and other factors, we adopted a systematic approach that identifies a relatively small number (2--10) of separate states of the detector for each segment and HV combination. \par

For this approach we used the \texttt{MySQL} database described in Section \ref{sec:obs}, storing all dark counts extracted from the existing dark exposures observed over the years. This database includes time of arrival for each event, its corrected location on the detector ($x, y$), event PHA, detector segment and HV, and the name and proposal ID of the dataset that recorded each of these counts. For each HV and segment, we first filtered these photons by adopting the following criteria: (1) we included only events recorded within the pre-defined inner region (see Section \ref{sec:obs}), and (2) we selected only those dark counts with PHA values between 2 and 27 as \texttt{CalCOS} by default excludes anything outside of these values. With a final list of events per setting (i.e. HV and segment), we then sorted the counts by their time of arrival. We divided all of the selected events into suites of 100,000 counts, and identified separate data suites. The dark counts from multiple files were combined to create individual data suites. We note that using a fixed number of dark counts (100,000) per data suite guaranteed that we probed the spatial distribution of photons on the detector evenly and consistently in each data suite. \par
\begin{figure*}
   
          \begin{minipage}{0.49\linewidth}
   	  \includegraphics[scale=0.55]{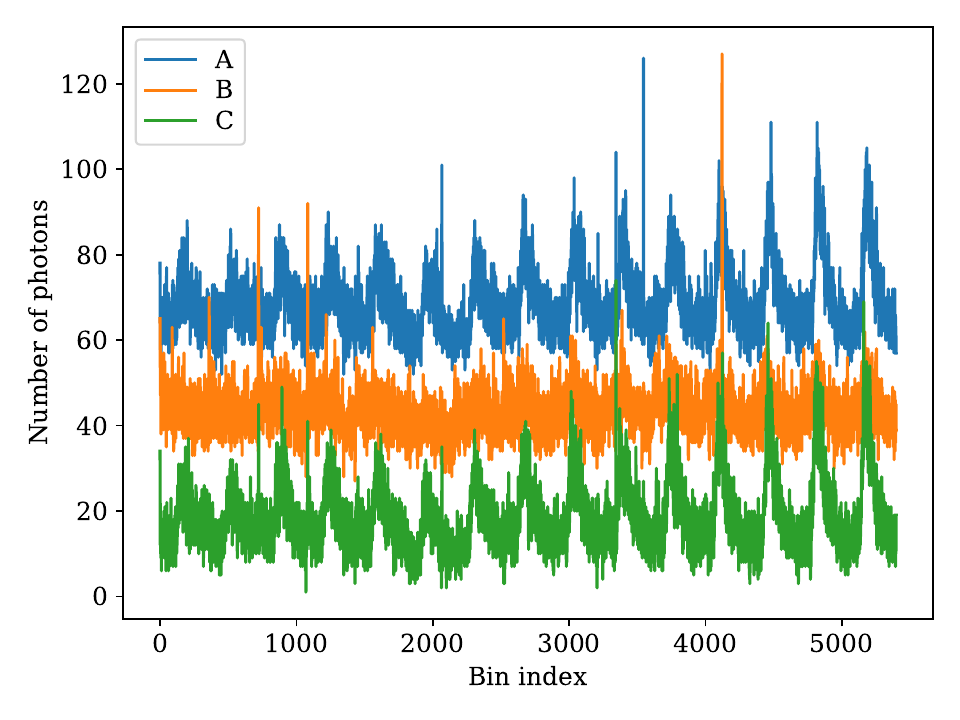}
   \end{minipage}
    \begin{minipage}{0.49\linewidth}
   	  \includegraphics[scale=0.55]{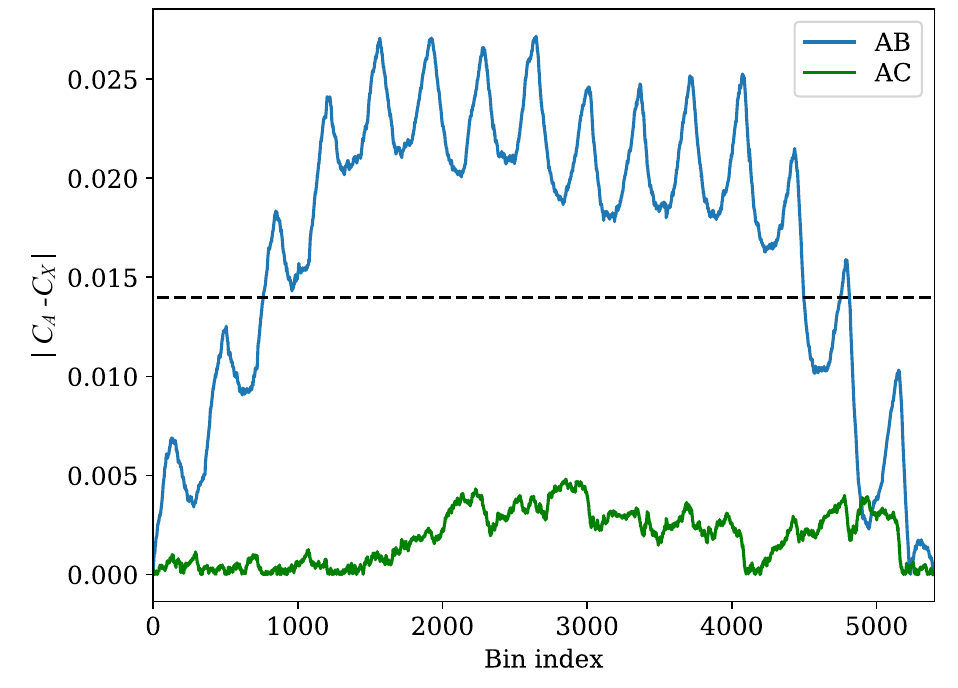}
   \end{minipage}
   \begin{minipage}{0.49\linewidth}
   	  \includegraphics[scale=0.55]{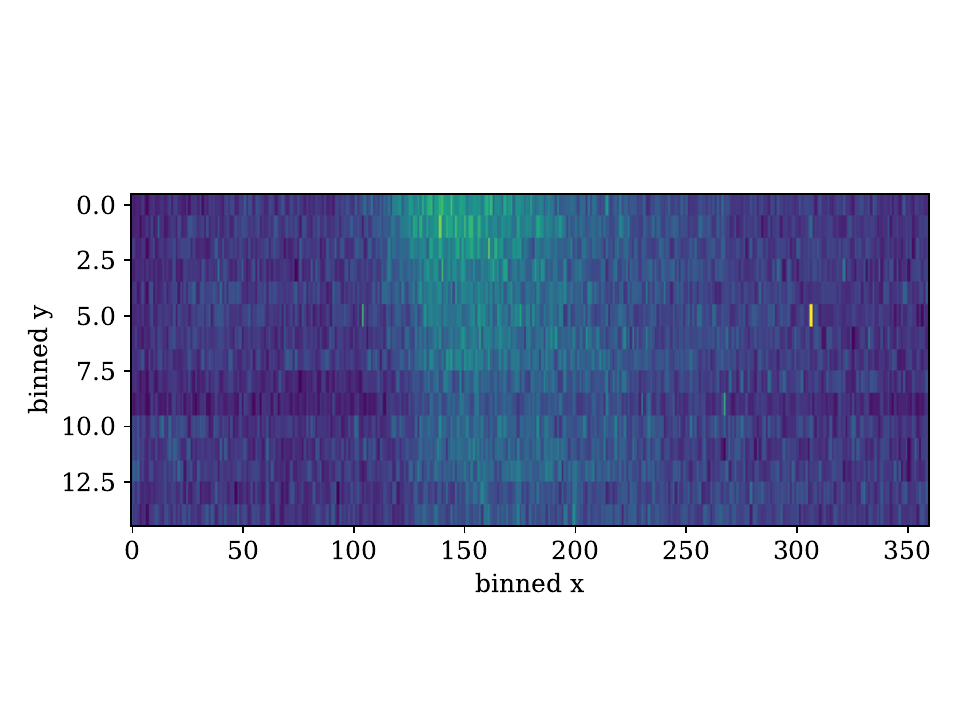}
   \end{minipage}
    \begin{minipage}{0.49\linewidth}
   	  \includegraphics[scale=0.55]{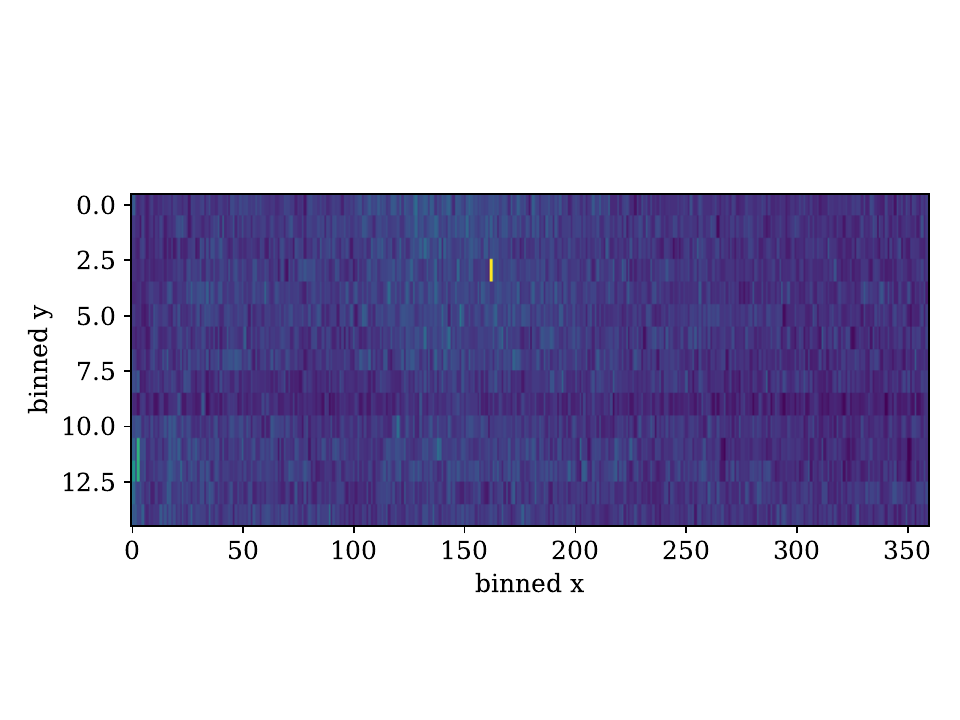}
   \end{minipage}

      \caption{Example for segment A, HV=167. Top left: reshaped data for data suites A, B and C. For visualization effects, B and C are shifted downward by 25 and 50 respectively in the vertical direction. Top right: values of $|C_A - C_X|$ as a function of data index. In our adopted approach we have chosen a  threshold of D $\geq$ 0.014, shown with a dashed line. Any data suites above this threshold are considered as a new detector state. On the bottom panels we show the 2D binned data suites for two different detector states identified by our approach. The bottom left shows data suite A and the bottom right shows data suite B. We highlight the present structure ("glow") in this particular state for segment A, which was entirely absent in any of the segment B states.}
         \label{fig:D_analysis}
   \end{figure*}

For each data suite we created 2D arrays which were binned into 360 bins in the $x$-direction and 15 bins in the $y$-direction. We reshaped these arrays into 1-dimensional arrays with length $360\cdot 15 = 5400$ (see top left panel in Figure~\ref{fig:D_analysis}). At this stage we compared different data suites to each other applying a test similar to the Kolmogorov-Smirnov \citep{ros11} test. Namely, we computed the cumulative distribution of counts along the data axis (along the index ranging from 0 to 5399), and normalized them. To compare data suite A with data suite B we computed:
\begin{equation}\label{eq:d}
D = \max | C_A - C_B |     
\end{equation}
where $C_A$ and $C_B$ are normalized cumulative distributions. We illustrate this particular step in our analysis in the top right panel in Figure~\ref{fig:D_analysis} for segment A, HV=167. Every time a data suite was identified as different from the previous ones based on its estimated $D$ value, it was added to the basis vector. This identification was done by comparing each data suite against all elements of the basis vector and adding individual new data suites to the vector only if they were significantly different based on their $D$ value (equation \ref{eq:d}); for our analysis we choose $D = 0.014$ as the threshold for selecting new detector states (shown as a dashed line in the top right panel of Figure ~\ref{fig:D_analysis}). This particular threshold was chosen based on the original two states, quiescent and high-activity superdarks. Visually, we can clearly confirm that data suite A and C in the top left panel in Figure~\ref{fig:D_analysis} are relatively similar to each other, describing a similar detector state. Once we compute $D$ for these two data suites, it is clear that their $D$ estimate is below the established threshold.\par
On the other hand, comparing the data suites A and B shows clear differences in the top left panel in Figure~\ref{fig:D_analysis}. The inferred $D$ estimate for these sets clearly reaches a value higher than the established threshold of  $D = 0.014$ (see the maximum value in the blue curve in the top right panel of Figure ~\ref{fig:D_analysis}), indicating that the two states sampled by data suites A and B are drastically different. 
Overall, the bottom panels in Figure \ref{fig:D_analysis} show how our approach identifies detector states where the spatial distribution of counts is clearly different, in this particular case for data suites A and B. \par
Once we identified a final number of detector states, we then created superdarks for each of these states by coadding the corresponding dark exposures as tracked in each of the data suites. To avoid having individual exposures included in multiple data suites, we treated  single exposures as non-divisible units. Lastly, we note that three separate states for segment A, at three different HVs (167, 173 and 178), were ultimately discarded from our final list of characterized states due to their extremely low number of events; even after binning both spatially and across PHAs, $>$ 5\% of superpixels contained zero events. 
In Table \ref{tab:superdark_states} we show the final number of detector states identified by our approach for each detector configuration, HV and segment.




\begin{table}
\caption{Number of detector states identified as part of our analysis.}
\label{tab:superdark_states}
\centering 
\begin{tabular}{ccccc}
\hline \hline
   Segment & HV & Number of states & Number of exposures\\
\hline
    A&  163  &  3 & 608\\
    A&  167  &  10 & 1515 \\
    A&  169  &   5 & 479\\
    A&  173  &   3 & 252\\
    A&  178  &   3 & 65\\
    B&  163  &   3 & 1215\\
    B&  167  &   3 & 589\\ 
    B&  169  &   4 & 1155\\
    B&  175  &   2 & 624\\
\hline
\end{tabular}
\end{table}

\section{ACDC: Another COS Dark Correction}\label{sec:acdc}
The optimal background correction algorithm was incorporated as the main component in our new software, \texttt{ACDC} (Another COS Dark Correction). Additionally, with the adopted approach described in Section \ref{sec:multi_states}, we characterized different states of the detector for each configuration and created superdarks (or empirical models) for each of them.\par
\texttt{ACDC} was developed as a Python tool, easily installable and released for public use\footnote{\href{https://github.com/jotaylor/acdc-hst}{https://github.com/jotaylor/acdc-hst}}, along with the suite of superdarks. Given that the background correction is an intermediate step in the COS calibration pipeline, to be able to replace the nominal \texttt{CalCOS} background correction with the optimized background method, we took the \texttt{corrtag}\footnote{\texttt{corrtag} files are intermediate \texttt{CalCOS} products. These files are binary tables containing corrected event lists.} files and processed them with \texttt{ACDC}. \texttt{ACDC} was designed to (1) determine the best dark model as described in Section \ref{sec:correction}, (2) use this model to subtract the estimated background counts from the science image, and (3) finally run the remaining \texttt{CalCOS} calibration steps to produce \texttt{x1d} (extracted 1D spectra) products. We note that \texttt{ACDC} identifies individual best-model superdarks for every single input exposure. As a final step, users can then coadd all of the corrected \texttt{x1d} files (for a given configuration) into a single spectrum by calling the flux-weighted \texttt{coadd} function developed as part of the ULLYSES initiative \citep{rom20} and installed as part of \texttt{ACDC}. \par
A few extra considerations were taken to further improve the performance of \texttt{ACDC}. One of these considerations included the integration of sigma-clipping when creating each superdark best model, to avoid including pixels or detector regions affected by hot or bad pixels. 
Additionally, some of the resulting superdarks displayed obvious gain sagged regions, particularly at the LP1 locations. To avoid obtaining a biased/incorrect superdark model for a given science exposure, \texttt{ACDC} makes use of the  information in the GSAGTAB\footnote{This reference file provides the locations of rectangular regions for portions of the FUV detector that have very low pulse height amplitude, known as gain sagged regions.} reference table appropriate for the science exposures and identifies heavily sagged regions (or the absence of them) intrinsic to the science frame, as these need to be present (or removed) in the best model superdark. Before initiating the likelihood maximization to select the best combination of superdarks for the science data, the software integrates the information on the gain sagged regions from the science exposures and when needed, interpolates over heavily sagged regions to match the expected detector state as observed in the science frames. This component of the code is particularly critical for accurately estimating the background contributions to the science spectra taken at LP1, before the gain sagged regions developed. \par 

\begin{figure*}
   	  \centerline{\includegraphics[scale=0.38]{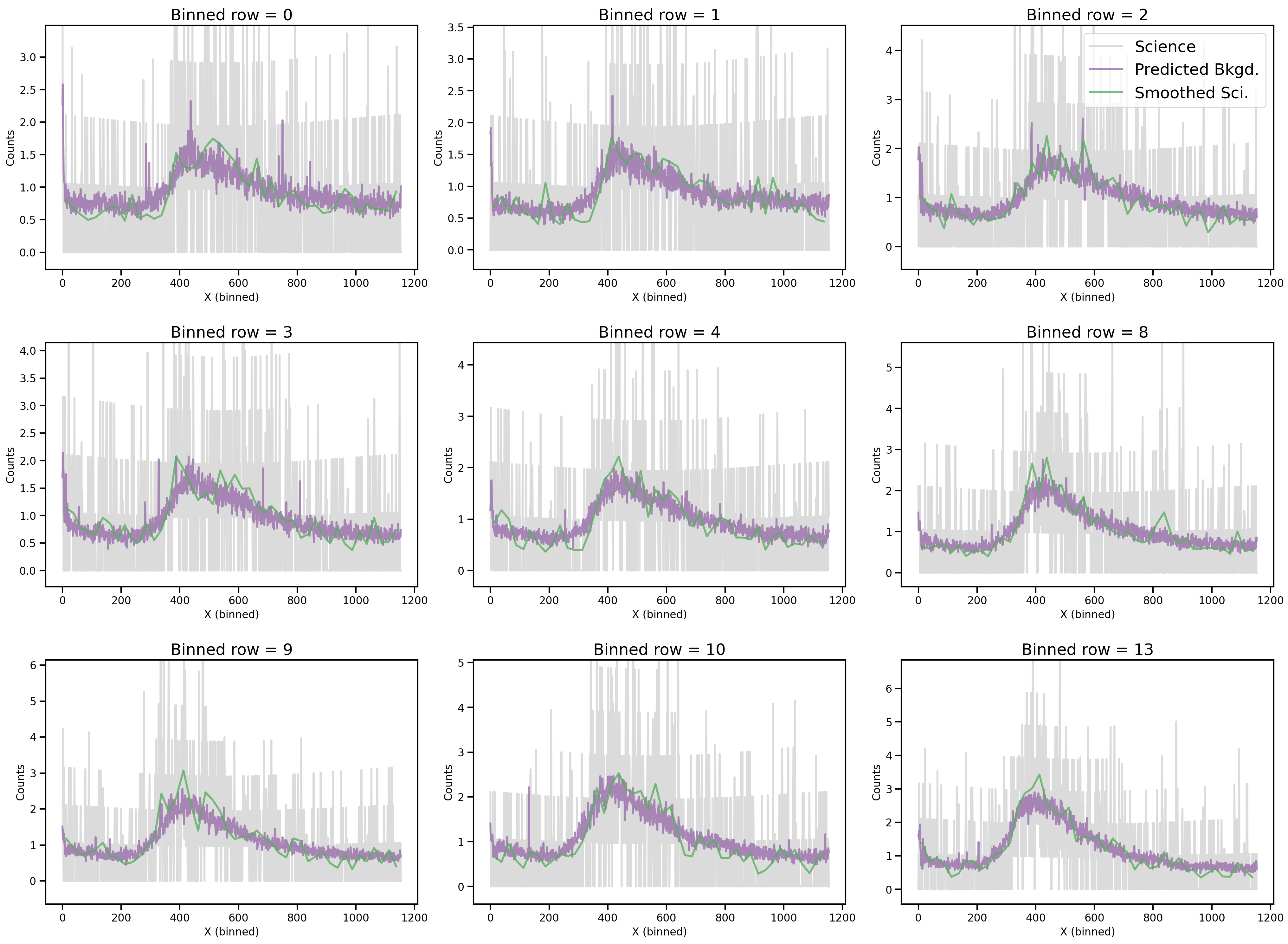}}
      \caption{Diagnostic plot for one of the datasets for target SDSS-J145735.13+223201.8 (LCAG02TPQ). We display the output from the optimized background correction code (\texttt{ACDC}) which uses the suite of multi-state superdarks. Subpanels show the dark count profiles in each binned row, excluding the science extraction and WCA regions. The dark profiles observed in the science exposures are shown in grey, the smoothed science exposures in green, and the best superdark model in purple. We highlight the agreement observed between the green and purple curves, showcasing the effectiveness of our adopted background correction approach.}
         \label{fig:acdc_2d_compare}
   \end{figure*}

Given the complexity of the optimized background correction in \texttt{ACDC}, to facilitate a quick inspection of the accuracy of the best superdark model, the software automatically generates diagnostic figures. These figures include plots of the dark profiles in the science exposures from each individual binned row (outside of the science and WCA regions) against those from the estimated best dark model. An example for one of the datasets (LCAG02TPQ) for target SDSS-J145735.13+223201.8, specifically for segment A, is shown in Figure \ref{fig:acdc_2d_compare}. The science extraction region for this exposure taken at lifetime position 5 is located in binned rows 6--8, and the WCA region is found in binned rows 11--12. Each subpanel shows, for each binned row: the counts in the \texttt{corrtag} file (grey), the mean binned \texttt{corrtag} counts over 25 superpixels (green), and the dark counts based on the best model superdark (purple). Overall, this figure highlights the accuracy of \texttt{ACDC} when creating the best superdark model, which matches the spatial dark variations observed in the 2D science exposure. Particularly notable is the presence of the "glow" feature seen only in segment A superdarks.

\subsection{ACDC vs. CalCOS}\label{sec:comparison}
\begin{figure*}
   	  \centerline{\includegraphics[scale=0.6]{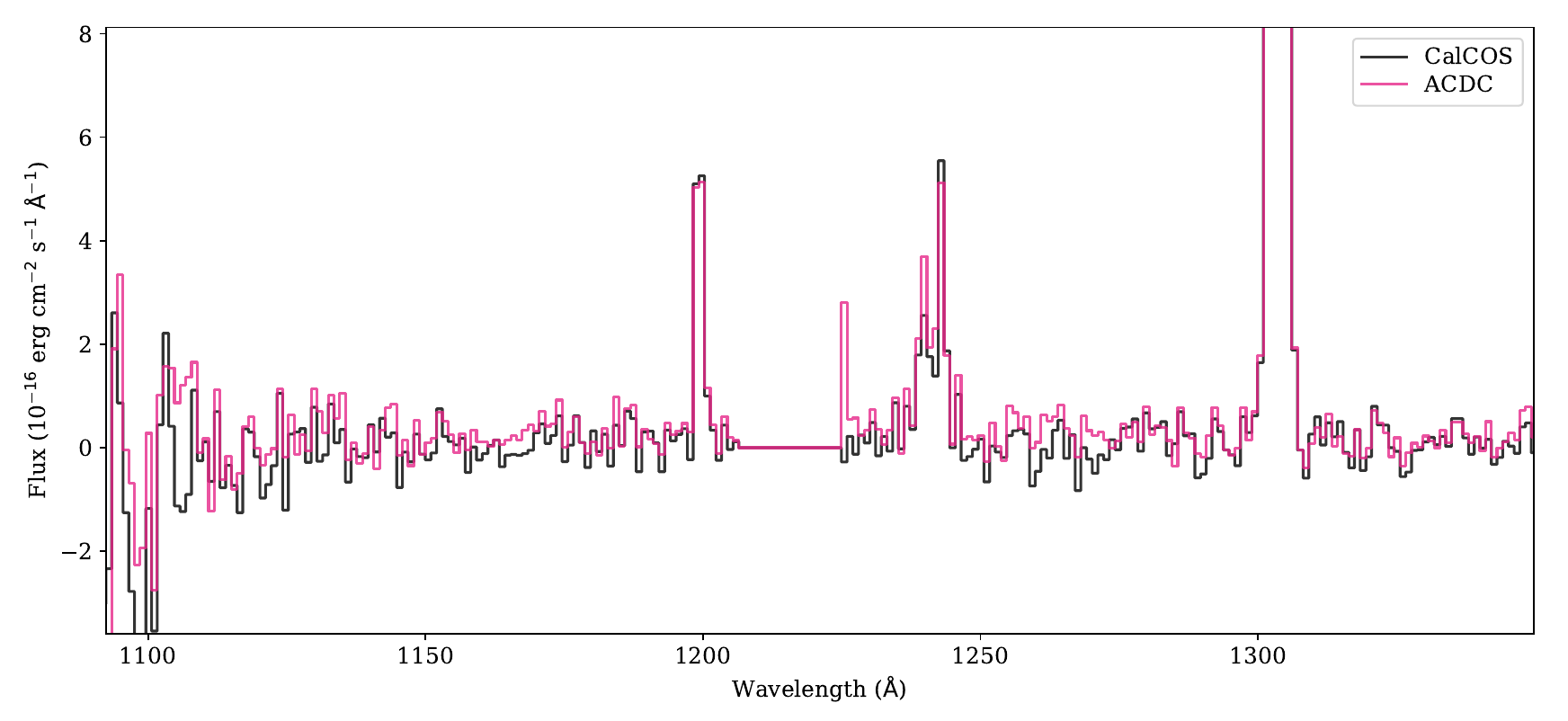}}
      \caption{HCG92-7 (PID: 13321, LCAY06020) spectra, taken with the COS FUV G130M/1222 setting, binned to 15 resolution elements for better visualization of the flux differences. The dataset was taken on August 13 2014. In black we show the standard \texttt{CalCOS} product; in magenta we show the \texttt{ACDC} product.  }
         \label{fig:acdc_hcg}
   \end{figure*}

\begin{figure*}
   	  \centerline{\includegraphics[scale=0.65]{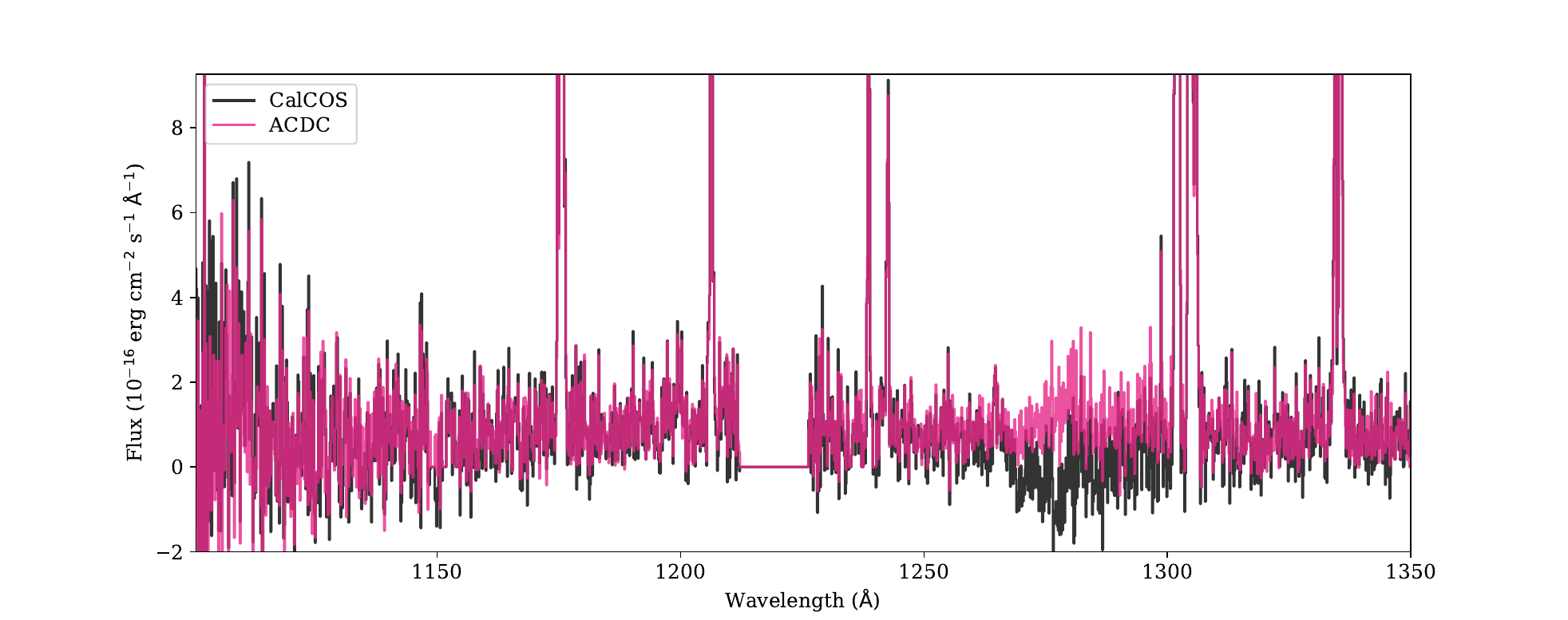}}
      \caption{2MASS-J15570234-1950419 (PID: 15310, LDMP24010) spectra, taken with the COS FUV G130M/1222 setting, binned to 2 resolution elements (12 native pixels). The dataset was taken on August 02 2018. In black we show the standard \texttt{CalCOS} product; in magenta we show the \texttt{ACDC} product. We highlight the over-subtraction in the \texttt{CalCOS} spectrum around 1275 \r{A}. }
         \label{fig:2mass}
   \end{figure*}

\begin{figure*}
   	  \centerline{\includegraphics[scale=0.25]{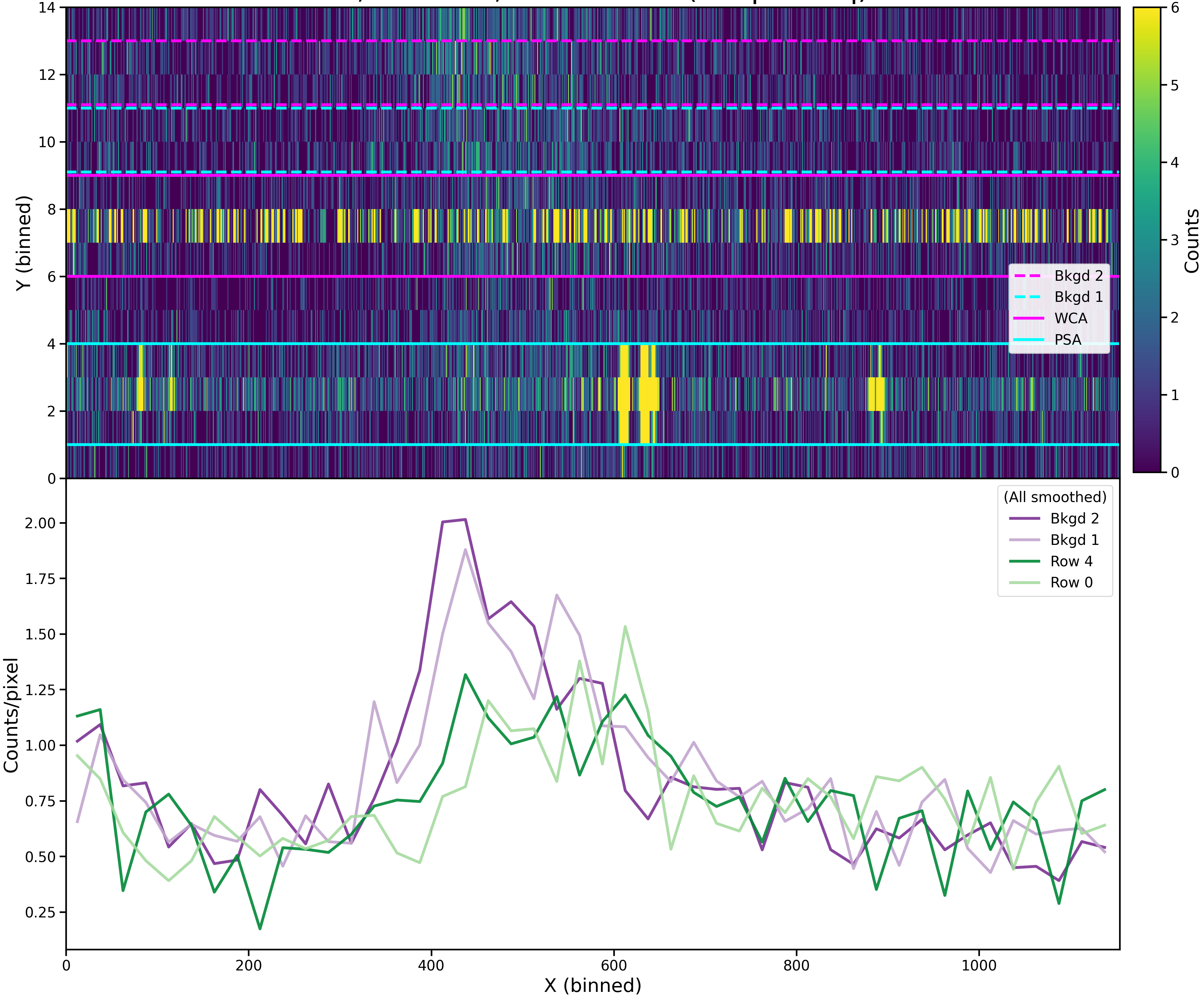}}
      \caption{\textit{Top:} 2D binned image of one of the science exposures (LDMP24UVQ) for target 2MASS-J15570234-1950419, taken at lifetime position 4 using segment A. The location of the science extraction region is shown with cyan solid lines. The location of the WCA is shown with magenta solid lines. The pre-defined background regions used by \texttt{CalCOS} to remove the background contributions to the science spectrum are shown with dashed lines, on the upper part of the detector. \textit{Bottom:} Smoothed 1D profiles for the two pre-defined background regions on the upper part of the detector, compared to the profiles of the rows below and above the science extraction region (binned rows 0 and 4). Note the drastic difference between the profiles.  }
         \label{fig:2mass_2d}
   \end{figure*}

We tested the new software and superdarks on a suite of low S/N observations (Table \ref{tab:test_suite}). In general, we find overall improvements performing a 2D background correction over the standard \texttt{CalCOS} pipeline (Figure \ref{fig:acdc_hcg}). These improvements are more accentuated in certain cases, specifically for segment A exposures that exhibit the presence of a glow-like structure (lower left panel in Figure \ref{fig:D_analysis}). To highlight this point, we show in Figure \ref{fig:2mass} the comparison of the calibrated products using  \texttt{CalCOS} (in black) against the data corrected using \texttt{ACDC} (in magenta) for target 2MASS-J15570234-19504. This particular dataset, LDMP24010, was taken as part of PID 15310 using the G130M/1222 setting on August 02, 2018. We confirmed that the over-subtraction observed in the final calibrated \texttt{CalCOS} product originates from an overestimate in the background contribution. The observations were taken at lifetime position 4, where the science spectrum falls at $y\sim$425, on the lower part of the detector (binned row 2; see Figure \ref{fig:2mass_2d}). The pre-defined background regions are instead located at $y\sim$ 565 and 630, on the upper part of segment A, exactly at the location of the``glow" (binned rows 9-12). The bottom panel in Figure \ref{fig:2mass_2d} clearly highlights the differences in the 1D profiles of the background on the upper part of the detector compared to those closer to the science extraction region towards the bottom of the detector. Our improved software, along with the multi-state superdarks, more accurately account and correct for the spatial background contributions at the location of the science spectrum. \par
We also note that part of the testing and validation of our software involved the inspection of the final S/N values for the two different calibration methods: \texttt{ACDC} and the default \texttt{CalCOS} calibration. Using the 2D optimized background correction approach, we found that on average the S/N values in the final 1D spectra are increased by $\sim$10\% on both segments compared to the exposures calibrated with the nominal \texttt{CalCOS} pipeline. However, we highlight that in a few cases the improvements in the S/N were as high as $\sim$ 60\%, both in localized wavelength regions (Figure \ref{fig:2mass}), or across the full wavelength range of individual segments.

\section{Conclusion}\label{sec:conclusion}
In this work we investigated methods for better characterizing the background on the COS FUV segments in order to perform a more-accurate dark correction, which is particularly critical for observations of faint targets. We carefully analyzed all dark exposures available in the STScI MAST archive as of September 2023, and confirmed that the distribution of dark counts on the COS detector varies spatially with time. We adopted a systematic approach that identifies between 2 and 10 different detector states per setting (segment and HV). We then created superdarks for each state and used these superdarks to perform a 2D background correction, accurately predicting and subtracting the expected dark counts in a given COS FUV science exposure. A much more detailed technical report describing our investigation into the background properties of the instrument can be found in our official documentation page\footnote{\href{https://github.com/jotaylor/acdc-hst}{https://github.com/jotaylor/acdc-hst}}. \par
Our optimal background correction algorithm is incorporated and released as the Python tool, \texttt{ACDC}\footnote{\href{https://github.com/jotaylor/acdc-hst}{https://github.com/jotaylor/acdc-hst}}, also available in Python Package Index (PyPI\footnote{\href{https://pypi.org/project/acdc-hst/}{https://pypi.org/project/acdc-hst/}}). This software, along with the publicly-available superdarks we have created, is able to perform a more accurate background correction than the default \texttt{CalCOS} calibration, specifically for low S/N datasets; our testing of \texttt{ACDC} showed that in several cases, the standard background correction incorrectly over-subtracted the dark counts due to the location of the static and pre-defined background regions. Additionally, we confirmed that in general \texttt{ACDC} increases the S/N of the final spectra by $\sim$10\% compared to that of the \texttt{CalCOS} products. We also note that implementing the approach adopted by \texttt{ACDC} into the standard \texttt{CalCOS} software would require extensive modifications to the flow of the pipeline\textbf{;} therefore for the time being \texttt{ACDC} is better used as a stand alone code. Overall, \texttt{ACDC} benefits COS FUV background-limited observations irrespective of the configuration, and promises to fully exploit the scientific potential of both archival and future data of faint targets. 

\acknowledgments
We thank the anonymous referee for helpful suggestions to improve the contents of this manuscript. The data analyzed in this paper were obtained from the Mikulski Archive at the Space Telescope Science Institute (MAST).



\bibliographystyle{aasjournal}
\bibliography{COS_Dark} 



\end{document}